\documentclass[showpacs,preprintnumbers,preprint,aps,amsmath,amssymb]{revtex4}
\usepackage{epsfig}
\def\lsim{\:\raisebox{-1.1ex}{$\stackrel{\textstyle<}{\sim}$}\:}
\def\gsim{\:\raisebox{-1.1ex}{$\stackrel{\textstyle>}{\sim}$}\:}
\def\10{$SO(10)$}
\def\21{SU(2) $\otimes$ U(1) }

\def\422{$SU(4) \otimes SU(2) \otimes SU(2)$}
\def\321{SU(3) $\otimes$ SU(2) $\otimes$ U(1)}
\def\lsim{\raise0.3ex\hbox{$\;<$\kern-0.75em\raise-1.1ex\hbox{$\sim\;$}}}
\def\gsim{\raise0.3ex\hbox{$\;>$\kern-0.75em\raise-1.1ex\hbox{$\sim\;$}}}
\def\vev#1{\left\langle #1\right\rangle}

\newcommand{\ba}{\begin{array}}
\newcommand{\ea}{\end{array}}
\newcommand{\be}{\begin{equation}}
\newcommand{\ee}{\end{equation}}
\newcommand{\beqa}{\begin{eqnarray}}
\newcommand{\eeqa}{\end{eqnarray}}
\relax
\def\321{$SU(3)\times SU(2)\times U(1)$}
\def\mt{$\mu$-$\tau$ }

\begin{document}
\bigskip
\title[]{Complex CKM matrix, spontaneous CP violation and generalized 
$\mu$-$\tau$ symmetry}
\author{
Anjan S.  Joshipura\footnote{anjan@prl.res.in} and Bhavik P. 
Kodrani\footnote{bhavik@prl.res.in}}
\affiliation{
Physical Research Laboratory, Navarangpura, Ahmedabad 380 009, India 
\vskip 2.0truecm}
\begin{abstract}
\vskip 1.0 truecm
The multi-Higgs models having spontaneous CP violation (SPCV) and natural 
flavor 
conservation (NFC) lead to a real CKM matrix $V$ contradicting current 
evidence 
in favour of a complex $V$. This contradiction can be removed by using a 
generalized $\mu$-$\tau$ (called 23) symmetry in place of the discrete 
symmetry 
conventionally 
used to obtain NFC. If 23 symmetry is exact then the Higgs induced 
flavour changing neutral currents (FCNC) 
vanish as in case of 
NFC. $23$ breaking introduces SPCV, a phase in $V$ and 
suppressed FCNC among quarks. 
The FCNC couplings $F_{ij}^{d,u}$ between $i$ and $j$ generations show a 
hierarchy
$|F_{12}^{d,u}|<|F_{13}^{d,u}|<|F_{23}^{d,u}|$  with the result that the 
FCNC can have 
observable consequences in $B$ mixing without conflicting with  the 
$K^0-\bar{K}^0$ mixing.
Detailed fits to the quark masses and the CKM matrix  are used to
obtain the (complex) couplings $F_{ij}^d$ and $F_{ij}^u$. Combined 
constraints from flavour and CP violations in the $K,B_d,B_s,D$ mesons are 
analyzed within the model. They allow ($i$) relatively 
light Higgs, 100-150 GeV ($ii$) measurable extra contributions to the 
magnitudes and phases of the $B^0_{d,s}-\bar{B}^0_{d,s}$ mixing amplitudes 
and 
($iii$) the $D^0-\bar{D}^0$ mixing at the current sensitivity level.
\end{abstract}
%\preprint{TIFR/TH/05-33}

\maketitle
 
%%%%%%%%%%%%%%%%%%%%%%%%%%%%%%%%%%%%%%%%%%%%%%%%%%%%%%%
%\section{Introduction}
%%%%%%%%%%%%%%%%%%%%%%%%%%%%%%%%%%%%%%%%%%%%%%%%%%%%%%%
While the exact source of the observed CP violation is still unknown, it 
is now clear \cite{cckm} that the Cabibbo Kobayashi 
Maskawa (CKM) matrix $V$ describing the charged weak interactions
is complex. This follows  from determination 
\cite{gam} of the angle   
$\gamma=-Arg(V_{ud}V_{cb}V_{cd}^*V_{ub}^*)$ using the tree decays of 
the type $B\rightarrow D~K$
which provide a clear evidence \cite{cckm} for a complex $V$ even if one 
allows
for arbitray new physics (NP) contribution to the loop induced processes 
in 
the standard model (SM). Explicit CP violation in SM automatically makes 
$V$ complex. In contrast, the standard  theories of 
spontaneous CP violation (SCPV) at the 
electroweak scale \cite{brbook} give a real CKM matrix.
These theories contain 
two or 
more Higgs doublets and lead to the Higgs induced   
flavour changing neutral current (FCNC). Discrete symmetry conventionally 
imposed \cite{brbook} in order to eliminate them automatically 
leads \cite{brreal} to a real $V$. 
Alternative possibility is not to eliminate  FCNC through a discrete symmetry 
but suppress them with a very heavy Higgs. This possibility also leads 
to
a suppressed CP phase and hence 
effectively a  real CKM matrix if SCPV occurs at the electroweak 
scale \cite{brm}. In either case, theories of SCPV need modifications
in order to accommodate a complex CKM matrix.
We propose here one such modification which also implies verifiable rich 
phenomenology.
 
Instead of eliminating FCNC altogether, we use 
a discrete symmetry  to selectively suppress them as was done in the past 
\cite{as}. 
Such selective suppression \cite{wein} 
may even be needed as there are several  arguments 
favoring new 
physics contributions in the $B-\bar{B}$ mixing \cite{np1,np2,bf} but not 
much 
in the $K$ system. Discrete symmetry we use is a generalization of the 
well-known \mt symmetry \cite{mt1} to the quark sector. This symmetry, 
when 
softly broken 
(1) leads to
spontaneous CP violation (2) can explain quark masses and entire complex 
CKM matrix  (3) gives rise to hierarchical FCNC with observable 
consequences in 
the $B-\bar{B}$ mixing. 

Consider the \21 model with two Higgs doublets $\phi_{a}$, ($a=1,2$) and a 
generalized \mt symmetry (to be called $23$ symmetry) acting on fermion 
fields as  $f_2\leftrightarrow 
f_3 $ with $\phi_2\rightarrow -\phi_2$. The Yukawa couplings are 
\begin{equation}
\label{yukawa}
 -{\cal L}= \bar{d}_L\Gamma^d_a\phi_a^0 
d_R+\bar{u}_L\Gamma^u_a\phi_a^{0*} u_R+ {\rm H.c.}~,
\end{equation}
CP
invariance makes $\Gamma^{u,d}_a$ real. Imposition of CP and the $23$ 
invariance
on the  scalar 
potential results in a  CP conserving minimum. We achieve SCPV here by 
allowing  soft breaking of  $23$ symmetry in the Higgs 
potential
through a term $\mu_{12} \phi_1^\dagger \phi_2$ whose presence 
along with other $23$ invariant terms violates CP spontaneously
\cite{brsoft}.  Without lose of generality we can assume 
$\vev{\phi_1^0}=v_1$ and 
$\vev{\phi_2^0}=v_2 e^{i\alpha}$. This leads to the
the quark mass matrices $M^q$ ($q=u,d$):
\begin{equation}\label{md} 
M^{d,u}=\Gamma_1^{d,u} v_1+\Gamma_2^{d,u} v_2 e^{\pm i 
\alpha}~,
\end{equation}
with
\begin{eqnarray}\label{gammas}
\!\!\!\!\Gamma_1^{q} v_1\!\equiv\!\left(\begin{array}{ccc}
X^{q}&A^q&A^q\\ 
A^q&B^q&C^q\\
A^q&C^q&B^q\\ \end{array}\right)\;\;\;
&\!\!\!\!\!\!\!\!,
\Gamma_2^q v_2\!\!\equiv\!\!\left(\begin{array}{ccc}
0&-A^q\epsilon_1^q&A^q\epsilon_1^q\\ 
-A^q\epsilon_1^q&-B^q\epsilon_2^q&0\\
A^q\epsilon_1^q&0&B^q\epsilon_2^q\\ \end{array}\right)~. \end{eqnarray}
In addition to imposing the $23$ symmetry, we have also assumed
that $M^q$ are symmetric as would be the case in $SO(10)$ with appropriate 
Higgs representations.

Eqs.(\ref{md}) and (\ref{gammas}) give the following key features of the 
model
\begin{itemize}
\item
The phase $\alpha$ in $M^q$ cannot be rotated away and leads to
a complex $V$. This can be seen by considering Jarlskog invariant 
$Det[M^uM^{u\dagger},M^dM^{d\dagger}]$ which is found to be non-zero 
as long as even one of $M^q$ is   
complex, i.e. $\epsilon_{1,2}^q\not=0$ for ($q=u$ or $d$). Thus unlike 
earlier 
models \cite{brbook}, a complex CKM originates here  from SCPV.
\item
An approximate  $23$ symmetry ($|\epsilon_{1,2}^q|\ll 1$) can explain 
\cite{mt2} the 
quark masses and mixing. $V_{ub}$ and $V_{cb}$ vanish in the 
symmetric limit and quark masses and the Cabibbo angle can be reporduced
with the hierarchy \cite{mt2}
\be \label{hier}
|X^q|\ll |\sqrt{2}|A^q|\ll |B^q|\sim |C^q|\approx 
\frac{m_3^q}{2} ~.\ee 
Small but non-zero  $\epsilon_{1,2}^q $ generate  
$V_{cb}$ , $V_{ub}$ and CP violation. Let
\begin{equation}
V^{q\dagger}_LM^qV^{q}_R=D^q~, \end{equation}
where $D^q$ is the diagonal mass matrix with real and positive masses.
The CKM matrix is given by $V\equiv V^{u\dagger}_L V^d_L$ and
phases in $V^q_{L,R}$ are chosen in such a way that $V$ has the standard
form advocated in \cite{pdg}.
In the  simplified case of CP conservation ($\alpha=0$) and
$\epsilon_1^q=0$ one finds \cite{mt2}, 
\begin{equation}
\label{vq} V^q_L=V_R^{q}= R_{23}(\pi/4) R_{23}(\theta_{23}^q)
R_{13}(\theta_{13}^q)R_{12}(\theta_{12}^q)~,\ee with
\be
\label{quarkmixing} \theta_{23}^q \approx 
-\frac{\epsilon_{2}^q}{2}~,
\theta_{12}^q\approx ~\sqrt{\frac{-m_{1q}}{m_{2q}}}~,
\theta_{13}^q\approx ~\frac{m_{2q}}{m_{3q}}\theta_{12}^q\theta_{23}^q~. 
\ee
giving $ V_{cb}\approx \frac{\epsilon_2^u-\epsilon_2^d}{2}~,~
V_{ub}\approx \theta_{12u} V_{cb} $. Thus $23$ breaking through 
$\epsilon_2^q$ not only generates $V_{cb}$ and $V_{ub}$ but also 
leads to relative hierarchy between them.
\item 
Like other 2 Higgs doublet models, eq.(\ref{yukawa}) generates FCNC but 
they are 
linked here to  $23$ breaking which also generates $V_{ub},V_{cb}$.
Both remain small if $23$ breaking is small.
Eqs.(\ref{yukawa},\ref{gammas}) can 
be manipulated to obtain
\be \label{fcnc}
-{\cal L}_{FCNC}=\frac{(2 \sqrt{2} G_F)^{1/2}m_{b}}{\sin\theta\cos\theta} 
F^d_{ij}\bar{d}_{iL} d_{jR}\phi_H
+{\rm H.C.} ~, \ee
where $\phi_H\equiv \cos\theta~\phi^0_2~e^{-i\alpha}-\sin\theta~\phi^0_1,
~~\tan\theta=\frac{v_2}{v_1}$ and 
\be \label{fij}
m_{b} F_{ij}^d\equiv (V^{d\dagger}_L\Gamma_2^d v_2 e^{i\alpha} 
V^{d}_R)_{ij}~,\ee
and we have introduced the physical third 
generation  quark mass $m_{b}$ 
as an overall normalization to make $F_{ij}^d$ dimensionless.
Analogous expressions hold in case of the up quarks.
Eqs.(\ref{gammas}) and (\ref{quarkmixing}) are used to show that
\beqa \label{nfij}
F_{12}^d&\approx& \frac{1}{2}\epsilon^d_{2}(\theta_{13}^d+2 
\theta_{12}^d\theta_{23}^d)\approx \pm 6.0\times 10^{-4}  ~,\nonumber \\  
F_{13}^d&\approx& \frac{1}{2}\epsilon^d_{2}\theta_{12}^d\approx 
\pm 7.0 \times 
10^{-3}~,\nonumber 
\\  
F_{23}^d&\approx& \frac{1}{2}\epsilon^d_{2}\approx \pm 4.0 \times 10^{-2} 
~,\eeqa
where the quoted numerical values 
follow  from the approximate  eqs.(\ref{quarkmixing}) by choosing 
$\epsilon_2^d\approx 
2 V_{cb}$.  
It is seen that  $F_{ij}^d$ are suppressed if 23 symmetry is mildly 
broken, $i.e.$ $|\epsilon_{1,2}^q|\ll1 $. Independent of this, they 
follow a hierarchy
\be \label{fhier}
 |F_{12}^d|\ll |F_{13}^d|\ll|F_{23}^d|~.\ee
Similar hierarchy holds in case of the up quarks also. This hierarchy is 
remarkable. The FCNC are suppressed most in the $K$ 
system where strong constraints on their existence already exist. In 
contrast, the flavour changing effects in the $B$ system can be more 
pronounced.  
\end{itemize}
The strength and hierarchy of $F_{ij}^q$ can be probed through
flavour changing transitions, particularly through $P^0-\bar{P}^0$, 
($P=K,B_d,B_s,D$) mixing. This  mixing is generated in the SM at 1-loop 
level and thus can become comparable to the tree level FC effects in 
spite 
of the suppression in $F_{ij}^q$. $P^0-\bar{P}^0$ mixing is induced by the 
element $M_{12}^P\equiv\langle P^0|{\cal H}_{eff}|\bar{P}^0\rangle$. The 
effective Hamiltonian here contains two terms ${\cal H}_{eff}^{SM}+{\cal 
H}^H_{eff}$ where the second term is induced from the Higgs exchange.
The ${\cal H}^H_{eff}$ follows from eq.(\ref{fcnc}) in a straightforward 
manner: 
\beqa \label{heff}
{\cal H}_{eff}^H(ij)&=&-\frac{2\sqrt{2}G_F m_b^2}{\sin^2 2\theta 
M_\alpha^2}(F_{ij}^{d^{2}}C_\alpha^2
(\bar{d}_{iL}d_{jR})^2+F_{ji}^{d*^{2}}C_\alpha^{*2}
(\bar{d}_{iR}d_{jL})^2\nonumber \\
&+&2 
F_{ij}^dF^{d*}_{ji}|C_\alpha|^2(\bar{d}_{iL}d_{jR})(\bar{d}_{iR}d_{jL}))~,\eeqa 
where  $ij=12,13,23$ respectively denote ${\cal H}_{eff}^H$ for the
$K,B_d,B_s$ mesons. The model contains three real Higgs fields $H_\alpha$
whose masses $M_\alpha$ appear above. The real and imaginary parts 
of $\sqrt{2}\phi_H\equiv R+i I$ in eq.(\ref{fcnc}) are related to 
$H_\alpha$ 
through a $3\times 3$ orthogonal matrix $O$ and one can write 
$\sqrt{2}\phi_H=(O_{R\alpha}+iO_{I\alpha}) H_\alpha\equiv C_\alpha 
H_\alpha ~$ 
which defines the complex parameters $C_\alpha$ appearing in 
eq.(\ref{heff}).

Define $F_{ij}^d\equiv|F_{ij}^d| e^{is_{ij}}, C_\alpha\equiv |C_\alpha| 
e^{i\eta_\alpha}$. Then using 
$|F_{ij}^q|=|F_{ji}^q|$ (following from symmetry of $M^q$) and the 
vacuum saturation approximation we obtain the Higgs 
contribution to $M_{12}(P)$ from eq.(\ref{heff}):
\be \label{mhp}
M^{H}_{12}(P)=\frac{\sqrt{2}G_F m_b^2 f_{P}^2 M_{P}|C_\alpha|^2 
|F_{ij}^d|^{2}}{6\sin^2 2\theta 
M_\alpha^2} Q_{ij}e^{i(s_{ij}-s_{ij})} \ee
with $$Q_{ij}=\left[A_P-1+10 A_P \sin^2 
(\frac{s_{ij}+s_{ji}}{2}+\eta_\alpha)\right]~$$
and
$A_P=\left(\frac{M_P}{m_a+m_b}\right)^2$, ($P^0\equiv \bar{a}b$).
$\Delta M^H(P)\equiv 2|M_{12}^H(P)|$ following from eq.(\ref{mhp}) depends 
on several unknown parameters 
in the Higgs sector while its phase is determined by the phases of 
$F_{ij}^d$ which depend only on parameters in $M^q$. For illustration, we 
retain the contribution of the
lightest Higgs $\alpha\equiv H$ in eq.(\ref{mhp}) and choose $M_H= 
150 GeV, \sin^2 2\theta=1, |C_H|^2=1/2, Q^P=1/2Q^P_{max}$. The 
numerical values of $F_{ij}^d$ in eq.(\ref{nfij}) then give
$$r^P\equiv |\frac{\Delta M^H(P)}{\Delta M^{exp}(P)}|\approx 
(0.25,0.26,0.11)$$
respectively for $P=B_d,B_s,K$.
It follows that effect of the FCNC can be suppressed in the model without 
fine tunning or without having very heavy Higgs. But they need not be 
negligible and can imply some new contributions which can be looked for.

The new physics contribution  to $M^{B_d,B_s}_{12}\equiv  
M^{d,s}_{12}$ has been 
parameterized in model independent studies \cite{np1,np2,bf} by
$$M^{d,s}_{12}=M^{d,s;SM}_{12} (1+\kappa^{d,s} e^{i\sigma^{d,s}}).$$
Unitarity of $V$, measurements of $|V_{us}|,|V_{cb}|,|V_{ub}|$ and the 
unitarity angle $\gamma$-all through tree level processes have been used 
to determine the allowed ranges in $\Delta M^{d,s}_{SM}\equiv 
2|M_{12}^{d,s;SM}|$. The hadronic 
matrix elements entering $\Delta M^{d,s}_{SM}$ are determined using 
lattice 
results and we will specifically use  predictions based on 
\cite{jlqcd}. The SM predictions along with the experimental determination 
of $\Delta M^{d,s}$ are used in \cite{bf} to obtain 
\be \label {smpredict}
\rho^d\equiv\left|\frac{\Delta M^d}{\Delta M^d_{SM}}\right|=
0.97\pm 0.39,\rho^s\equiv\left|\frac{\Delta M^s}{\Delta M^s_{SM}}\right|=
1.08\pm 0.19~.\end{equation}
Possible presence of new physics contribution is hinted by the phase
$\phi_d$ of $M_{12}^d$ : $\phi_d=43.4^\circ\pm 2.5^\circ$ which differs 
from
its value $53.4^\circ\pm 3.8^\circ$ in SM determined \cite{bf} using 
$|V_{ub}|$ measured in inclusive $b\rightarrow u l\nu$. This implies a
non-zero new physics contribution $\phi_d^{NP}=Arg(1+\kappa_d e^{i 
\sigma_d})=-(10.1\pm 4.6)^\circ$ which in the present case can come from 
the Higgs exchanges. 
The values of $\rho_{d,s}$ and 
$\phi_{d}$ have been used to determine allowed ranges in the parameters
$\kappa$ and $\sigma$. This is displayed in Fig.(1) in case of the $B_d$
mesons. We can confront these observations now with the specific 
predictions in the present case.

Our strategy is to first determine parameters in $M^q$ from the quark 
masses and mixing and then use them to determine $F_{ij}^q$ which are used 
to obtain information on $M_{12}^H(P)$. Since the number of parameters in 
$M^q$ is more than the observables, we do our analysis in two ways.
First, we allow all parameters in $M^q$ to be free and determine them by
minimizing 
$\chi^2$:
$$\chi^2=\sum_{i=1,10}\left(\frac{E_i(x)-\bar{E}_i}{\delta 
E_i}\right)^2~,$$
where $E_i(x)$ represent predictions of six quark masses, three moduli
$|V_{us}|,|V_{cb}|,|V_{ub}|$ and the Jarlskog invariant $J$ calculated as 
functions of parameters of $M^q$.
The quantities $\bar{E}_i\pm \delta E_i$ are their values determined from
experiments. We choose quark masses at $M_Z$ given in \cite{quarkmasses} 
and 
all the 
CKM elements except $|V_{ub}|$ as in \cite{pdg}. For the latter, we use
the value $(4.4\pm0.3)\cdot 10^{-3}$ based on the determination \cite{bf} 
from the inclusive $b$ decays. We find many 
solutions giving excellent fits with 
$\chi^2\lsim 10^{-7}$. One specific example is 
given in the table. The parameters of the table lead to
\begin{equation}
\begin{array}{ll}
F_{12}^d=~~(0.26 -1.19 ~i)\cdot 10^{-4},&
F_{21}^d=~~(0.096 +1.22 ~i)\cdot 10^{-4};\\
F_{13}^d=-(0.53+5.2~i)\cdot 10^{-3},&
F_{31}^d=-(5.1+0.85~i)\cdot 10^{-3};\\
F_{23}^d=~~(0.30+1.13~i)\cdot 10^{-2},&
F_{32}^d=-(1.11+0.35~i)\cdot 10^{-2};\\
F_{12}^u=~~(-2.1+1.3~i)\cdot   10^{-4},&
F_{21}^u=-(1.3+2.1 ~i)\cdot 10^{-4} ~.\\ 
\end{array} \end{equation}
which are similar to the rough estimates in eq.(\ref{nfij}). 
\begin{table}
%\centering
\begin{math}
\begin{array}{|c|c|c|c|c|c|c|} \hline
&X~({\rm GeV})&A~({\rm 
GeV})&B~({\rm GeV})&C~({\rm GeV})&\epsilon_1&\epsilon_2\\
\hline
{\rm up}&0.0019&0.036&90.66&-90.04&-0.198&-0.082\\
\hline
{\rm down}&0.0035&0.019&1.54&-1.46&0.177&-0.025\\
\hline
\end{array}
\end{math}
\caption{An example of fit to the quark masses and mixing angles
corresponding to $\chi^2=1.4\cdot 10^{-7}$ and $\alpha$=-3.899.} 
%\label{tab:Multiplet}
\end{table} 

The above fits strongly depend on some of the 
$\epsilon_{1,2}^q$ being non-zero since if they vanish then  
$|V_{ub}|,|V_{cb}|$ and CP 
violation also vanish. However, we could get 
excellent fits with $|\epsilon_{1,2}^q|< 
0.2$ showing that approximately broken $23$ symmetry provides a very good 
description of the quark spectrum.
In an alternative analysis, we fixed $B^q,C^q$ from 
$|B^q-C^q|=m_{3q},|B^q+C^q|=m_{2q}$ which correspond to $23$ symmetric 
limit in the two generation case. This limit is found to be quite
good and gives good fits with $\chi^2\lsim 1$ when it is minimized 
with respect to the remaining nine parameters.

The predictivity of the scheme comes from the fact that each set 
of parameters of $M^q$
determined as above completely fix (complex) FCNC strengths $F_{ij}^{u,d}$
-in all 18 independent real quantities. We use these predicted values
to calculate Higgs 
contribution to $M_{12}^P$ by randomly varying unknown parameters of 
eq.(\ref{mhp}). We 
retain 
contribution of only one Higgs and vary its mass from 100-500 GeV.
$|C_\alpha|~,~\sin^2 2\theta$ and 
the phase $\eta_\alpha$ are varied over their full range namely,
$0-1$ and $0-2 \pi$ respectively. We neglect the possible 
corrections to the vacuum saturation approximation but 
vary $f_{B_{d,s}}$ over the full $1\sigma$ range:
$f_{B_s}=0.24\pm 0.04$ GeV, $f_{B_d}=0.2\pm 0.025$ GeV.
We require that ($i$) $\rho_{d,s}$ and 
$\phi_{d}$ lie in the allowed $1\sigma$ range ($ii$) The Higgs 
contribution 
to the $D^0-\bar{D}^0$ mixing amplitude satisfy the bound $|M_{12}^{DH}|< 2.2 
\cdot 10^{-14}$ GeV derived in \cite{nir} from  the BaBar and Belle 
measurements ($iii$) 
the Higgs contribution to the $K^0-\bar{K}^0$ mass difference and to 
$\epsilon$ is an order of magnitude less than their central experimental 
values. Combined results of this analysis for several 
sets of allowed
$F_{ij}^q$ are shown as scattered plot in Fig.(1). 

\begin{figure}[ht]
\begin{center}
\includegraphics[height=7.6cm,width=0.45\textwidth,angle=-90]{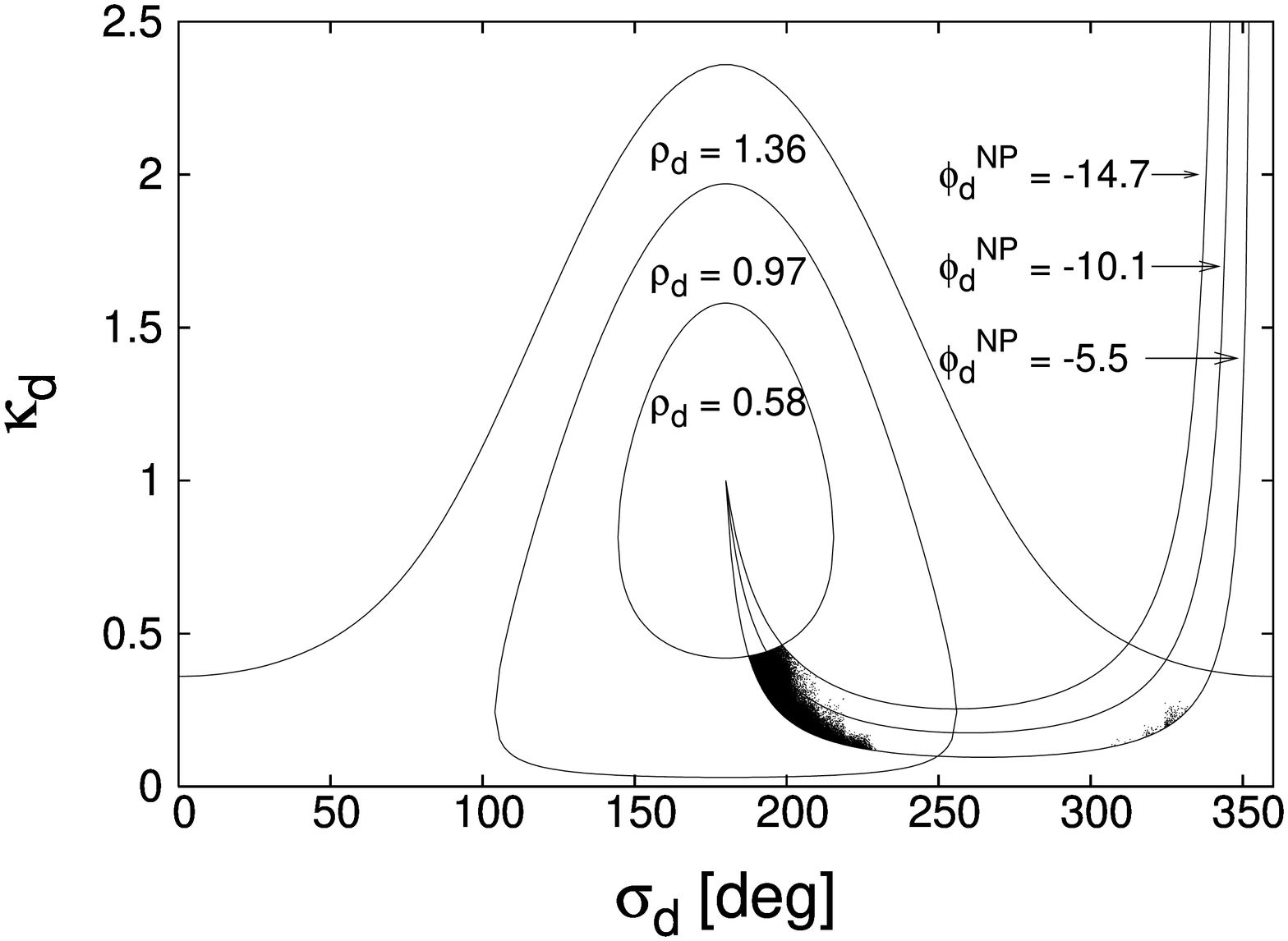}\hskip 
0.75cm 
\includegraphics[height=7.6cm,width=.45\textwidth,angle=-90]{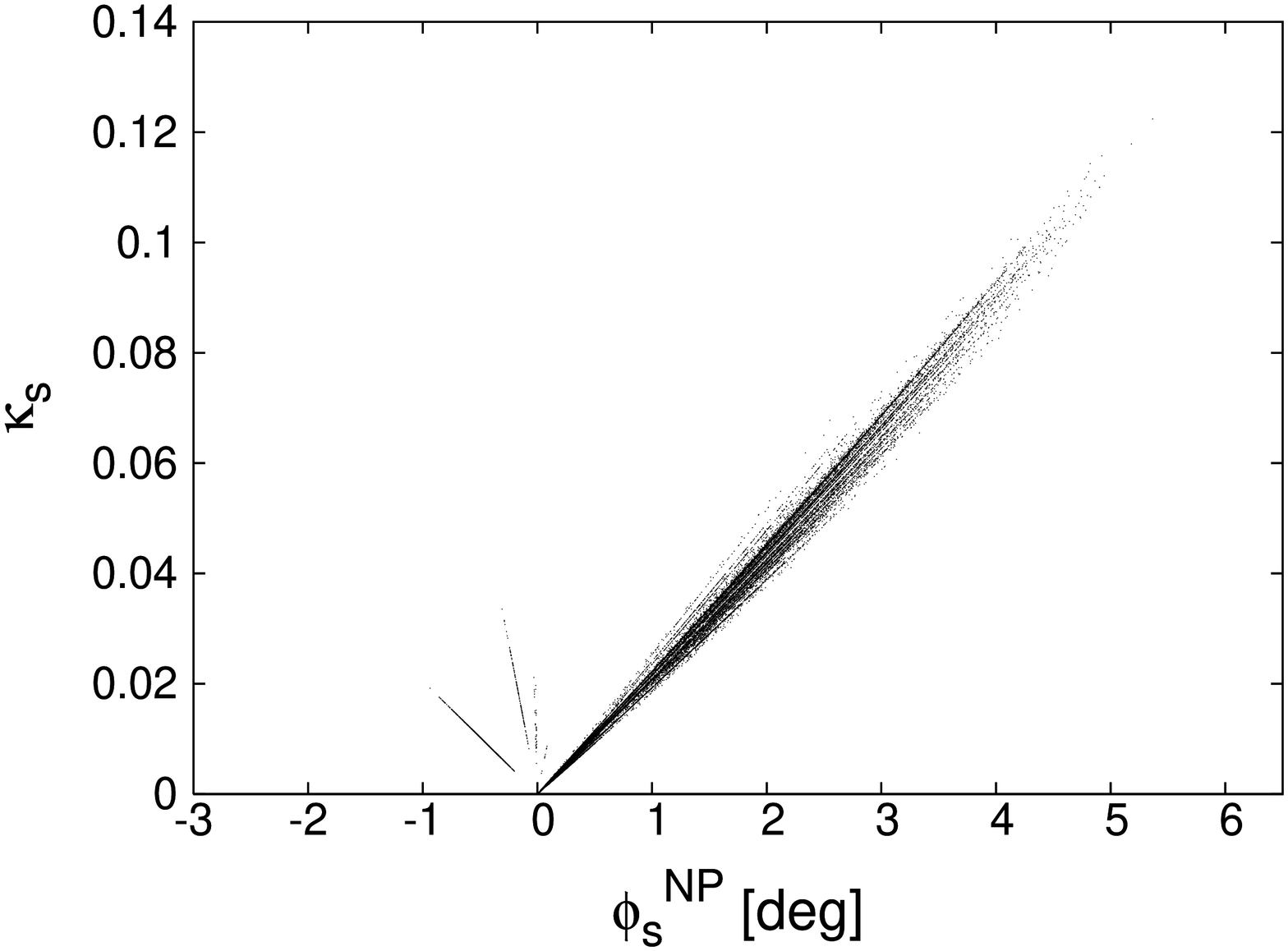}
\end{center}
\caption{Allowed regions in the $\kappa_d,\sigma_d$ (left panel) and 
$\kappa_s,\phi_s^{NP}$ (right panel) planes  
following from the inclusive determination of $|V_{ub}|$ and JLQCD 
result for the hadronic matrix element. Solid lines in the left 
panel corresponds to 
$1\sigma$  allowed values for $\rho_d,\phi_d^{NP}$ in model independent
study. The scattered plots in both panels
correspond to the predictions of the present model.} 
\end{figure}

The solid curves describe restrictions on $\kappa_d,\sigma_d$ following 
from eq.(\ref{smpredict}) and the measured value of $\phi_d$ in a model 
independent study. In the present case, 
the allowed values of $\kappa_d,\sigma_d$ are indirectly effected by 
restrictions coming from mixing of other mesons as well since the same set 
of Higgs parameters determine these mixings. Thus 
simultaneous imposition of the above mentioned constraints considerably 
restrict the allowed ranges in parameter space shown as scattered plot in 
Fig.(1). $\sigma_d$ is restricted 
in such a way that the Higgs contributes negatively to $\rho_d$ (in 
most parameter space) and  reduces the 
value of $\rho_d$ compared to the SM. $\kappa_d$ and 
$\sigma_d$ are restricted 
in the range $0.2<\kappa_d<0.46~, 185^\circ\lsim\sigma_d\lsim 229^\circ$ 
which 
correspond to $0.58\lsim\rho_d\lsim 0.9$ and $\phi_d^{NP}\approx 
-(5-15)^\circ$. 

The right panel in Fig.(1) shows the predictions of $\kappa_s$ and 
possible new physics phase $\phi_s^{NP}\equiv Arg(1+\kappa_s e^{i 
\sigma_s})$ in $M_{12}^s$. The allowed values of $\kappa_s$ after the 
combined constraints from all sources are relatively small $\leq 0.1$. 
This also 
results in
a small $\phi_s^{NP}$ although the Higgs induced CP phase $\sigma_s$ could 
be large. $\phi_s^{NP}$ may be approximately identified with the CP 
violating phase $\phi_s$ in the semileptonic CP asymmetry of $B_s$ decay. 
The maximum allowed value $\phi_s^{NP}\approx \phi_s\approx 0.1$ in the 
model is 
much larger than the SM contribution ($\phi_s\approx (4.1\pm 
1.4)\cdot 10^{-3}$) and is consistent with the present value \cite{d0}  
$|\phi_s|=0.70 ^{+0.47}_{-0.39}$. Possible improvement in the value of 
$\phi_s$  at LHC  would provide a crucial test of the model.

It is found 
that the $D^0-\bar{D}^0$ mixing plays an important role in ruling out some 
of 
the regions in parameter space and in most of the allowed regions 
$|M_{12}^{DH}|$ remains close to the limit $2.2 \cdot 10^{-14}$ GeV.
 
The above considerations used the $|V_{ub}|$ 
determined from the inclusive
$b\rightarrow u l\nu$ decay. We have repeated the analysis using the 
corresponding result
$|V_{ub}|=(3.8\pm 0.6)\cdot 10^{-3}$ from the exclusive decay.
We find that
regions in parameter space get shifted compared to Fig 1.  

In summary, we have addressed the problem \cite{alt} of obtaining a 
phenomenological 
consistent picture of SPCV. This is an important issue in view of the fact 
the earlier theories of SCPV led to a real CKM matrix while recent
observations need it to be complex. The proposed picture is 
phenomenologically consistent and does not need very heavy Higgs to 
suppress FCNC present in general multi Higgs models. The hierarchy in FCNC
, eq.(\ref{fhier}) obtained here through a discrete symmetry has 
observable 
consequences. The effect of Higgs is to reduce the
$B^0_d-\bar{B}^0_d$ mixing amplitude compared to 
the standard model prediction. The $D^0-\bar{D}^0$ mixing can be 
close 
to the bound derived from observation \cite{nir}. The  
new contribution to the magnitude of  $B^0_s-\bar{B}^0_s$ mixing is
small. The Higgs induced phase in this mixing  is found to be relatively 
low but much larger than in the SM.

Noteworthy feature of the proposal is universality of the discrete 
symmetry used here. The generalized \mt symmetry used here can 
explain large atmospheric mixing angle in the manner proposed in 
\cite{mt2} on one hand and can also account for the 
desirable features of the quark mixing and CP violation as discussed 
here. Details of  a unified description and constraints from other
flavour and CP violating observables will be 
discussed elsewhere. 

\end{document}